\documentclass[runningheads]{llncs}

\usepackage[T1]{fontenc}

\usepackage{graphicx}

\usepackage{hyperref}

\usepackage{booktabs}
\usepackage{multirow}
\usepackage{float}
\usepackage{subcaption}
\usepackage[dvipsnames]{xcolor}
\usepackage{amsmath}
\usepackage{makecell}

\usepackage[most]{tcolorbox}
\tcbuselibrary{skins, breakable}
\usepackage{adjustbox}

\begin{document}

\title{Can LLMs Extract Architectural Design Decisions from Source Code Commits? -- \newline 
A Preliminary Exploratory Study}

\titlerunning{Extracting ADDs from Code Commits using LLMs: Preliminary Study}

\author{Amey Karan\inst{1} \and
Rudra Dhar\inst{1} \and
Mohamed Soliman\inst{2} \and
Karthik Vaidhyanathan\inst{1}
}
\authorrunning{A. Karan et al.}

\institute{International Institute of Information Technology Hyderabad, India 
\\\email{\{amey.karan, rudra.dhar\}@research.iiit.ac.in, karthik.vaidhyanathan@iiit.ac.in}
\and
Paderborn University, Germany
\email{mohamed.soliman@uni-paderborn.de}
}

\maketitle              

\newcommand{\replpkg}{doi.org/10.5281/zenodo.20996340}

\begin{abstract}

\textbf{Context:}
Architectural Design Decisions (ADDs) capture the rationale behind the structure and evolution of software systems but are rarely documented explicitly, and are often hidden inside source code commits. Recovering them is important for Architectural Knowledge Management (AKM).

\textbf{Problem:}
Extracting ADDs from commits is challenging due to their implicit and unstructured nature. Large Language Models (LLMs) have shown strong capabilities in understanding code and text, yet their effectiveness for this task remains underexplored.

\textbf{Study:}
We present a preliminary study using four LLMs (Gemini~3 Pro, DeepSeek~R1, Kimi~K2, Qwen3) with zeroshot and fewshot prompting on 30 developer-written ADDs from open-source projects. We score outputs with ROUGE-L, BLEU, METEOR, and BERTScore, and one author manually reviews the Gemini outputs.

\textbf{Results:}
All models reach a BERT-F1 above 0.81, and fewshot prompting improves alignment (Gemini BERT-F1: $0.828 \rightarrow 0.847$). However, the generated ADDs are often too long, implementation-focused, and miss the rationale behind the decision. This highlights opportunities for architecture-aware LLM systems and automated AKM.

\end{abstract}

\section{Introduction}

Software architecture can be seen as a set of Architectural Design Decisions (ADDs) that shape a system's structure, behaviour, and evolution \cite{sa_set_design_decision}. ADDs are at the heart of Architectural Knowledge Management (AKM), but are rarely written down \cite{CAPILLA2016191,add_survey}. Instead, architectural information is spread across artifacts like issue trackers, mailing lists, blogs, and source code commits \cite{soliman2016exploring,soliman2017where,Bhat,Shahbazian_Medvidovic}.

Existing extraction approaches mostly rely on rich textual artifacts such as issue trackers and mailing lists, which are often incomplete or missing, leaving much architectural knowledge unrecovered \cite{add_survey,Bhat}.
Source code commits are the most consistently available artifact of how a system evolves and are known to carry architectural information \cite{soliman2016exploring,Bhat}.
However, commits primarily describe low-level implementation modifications, whereas ADDs are typically expressed at a higher level of abstraction. Therefore, recovering ADDs from commits requires identifying architecturally significant changes and interpreting their broader design intent and rationale.

Large Language Models (LLMs) work well on code and natural language \cite{feng2020codebert,chen2021codex,li2023starcoder}. This makes LLMs a promising candidate for extracting architectural knowledge from software repositories. However, it is not clear how well they can recover ADDs from commits, particularly when architectural rationale and context are only implicitly represented.

In this paper, we present a preliminary exploratory study on extracting ADDs from commits using LLMs. We compare zeroshot and fewshot prompting on four LLMs using 30 developer-written ADDs from open-source projects, yielding 60 generated decisions. These are scored using ROUGE-L, BLEU, METEOR, and BERTScore. To complement the metrics, an author also manually reviewed the outputs of the best model (Gemini~3 Pro) along three dimensions -- \emph{abstraction level}, \emph{verbosity}, and \emph{rationale}. All models reach a BERT-F1 above 0.81 and fewshot prompting improves alignment, but outputs are often too verbose, include implementation details, and leave out the rationale.
We provide a replication package with the dataset, prompts, and scripts\footnote{\href{https://\replpkg}{Replication package -- \replpkg}}.

\section{Related Work}
\label{sec:related_work}

Early works in automated AKM, such as Bhat et al.\ \cite{Bhat} and Shahbazian et al.\ \cite{Shahbazian_Medvidovic} show that issue discussions contain recoverable architectural knowledge, while tools like ADeX extract ADDs from project history artifacts \cite{Bhat}. However, these approaches depend on detailed textual documentation, which is often incomplete or unavailable in practice \cite{add_survey}.
More recent work explores LLM-based support for architectural decision-making. Systems such as \textit{ArchMind} and \textit{Archify} assist architects using explicit descriptions of context and alternatives \cite{archmind,marinho2021archifyrecommenderarchitecturaldesign}. Other studies evaluate whether LLMs can generate ADRs from rich contextual inputs, existing ADRs, or high-level descriptions \cite{soliman_llms4sa,amalfitano2025sar,dhar2024llm_adr}. While these studies show that LLMs can generate structured architectural artifacts, they also report issues such as inconsistency and inappropriate abstraction levels \cite{jasmin_state}.
To improve generation quality, later work focuses on prompting strategies and multi-stage pipelines. Maranh\~ao et al.\ \cite{maranhao_prompt} use structured prompt sequences, while Rodriguez-Sanchez et al.\ \cite{rodriguez_sanchez2025llm_ensemble} apply prompt and model ensembles for robustness. Other studies show that prompt design and contextual structure strongly affect ADR and rationale generation quality \cite{dhar2024llm_adr,zhou2025llm_dr}. However, these approaches still rely on explicit high-level textual context rather than raw commit history.

In contrast, recovering ADDs directly from source code commits remains underexplored. This creates a practical gap because commits are often the only consistently available record of architectural evolution. Our work addresses this gap by evaluating how effectively LLMs can recover ADDs from commits alone and how prompting and input structuring affect extraction quality.
\begin{figure}[t]
\centering
\includegraphics[width=\textwidth]{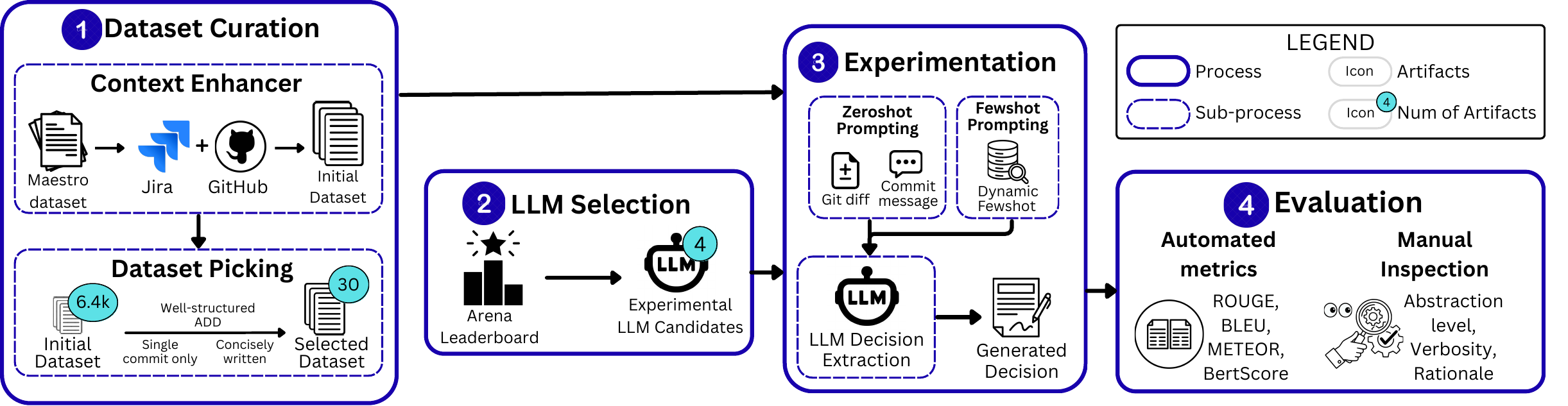}
\caption{Overview of Experimental Setup} \label{fig:study_diagram}
\end{figure}

\section{Study Design}
\label{sec:study_design}

We conduct a preliminary study using a small dataset of developer-written ADDs from open-source systems. 
We use the term \textit{alignment} to describe how closely a generated ADD reflects developer-documented ADD in terms of architectural intent, abstraction, and semantic meaning.

\subsection{Research Questions}

\noindent
\textbf{RQ$_1$}: \textit{To what extent do ADDs extracted by LLMs from source code commits using zeroshot prompting align with documented ADDs?} 
\noindent
\textbf{RQ$_2$}: \textit{How does fewshot prompting influence the alignment between extracted and documented ADDs?}


\begin{table}[b]
\caption{LLMs used in the study.}
\label{tab:llm_details}
\centering
\begin{adjustbox}{width=0.90\textwidth}
\begin{tabular*}{\textwidth}{@{\extracolsep{\fill}}lllll}
\toprule
\textbf{Task} & \textbf{Model} & \textbf{Parameters} & \textbf{Provider} & \textbf{License} \\
\midrule
\multirow{4}{*}{Experimental} 
& Gemini 3 Pro & Unavailable & Google & Proprietary \\
& DeepSeek R1 & 685B & DeepSeek AI & Open-source \\
& Kimi K2 Thinking & 1T (32B active) & Moonshot AI & Open-source \\
& Qwen3 & 32B & Alibaba Cloud & Open-source \\
\bottomrule
\end{tabular*}
\end{adjustbox}
\end{table}

\subsection{Experimental Setup}

\subsubsection{Dataset Curation}

We start from the Maestro dataset \cite{soliman_maestro}, which we choose as it already provides a mapping between Jira issues and the presence of ADDs, comprising 6,416 Jira tickets labelled for ADDs across 173 open-source projects. We link each ticket to its commit using the ticket ID found in commit messages, giving an initial dataset of $(\textit{commit}, \textit{ADD})$ pairs. From this, we manually picked 30 pairs that satisfy the following criteria: (1) the ADD is clearly written, (2) it maps to a single commit, and (3) the ADD description is between 60 and 100 words. These ADDs come from Apache Cassandra, Hadoop, and Tajo repositories, and cover technology choices, component-level changes, and quality-attribute changes. The dataset can be found in our \href{https://\replpkg}{replication package}.


\subsubsection{LLM Selection}

We pick four LLMs for this study, shown in Table~\ref{tab:llm_details}. The models are selected from the Arena Leaderboard\footnote{\href{https://arena.ai/leaderboard/text}{Arena Leaderboard; accessed 2026-01-03 -- arena.ai/leaderboard/text}}, a widely accepted LLM benchmark \cite{lmarena}. We choose the models based on their ranking, size, and  availability. Gemini~3 Pro is the top proprietary model, while DeepSeek~R1 and Kimi~K2 are the two largest open-source models. Qwen3 is the smallest model we can run locally.


\subsubsection{Experimentation}
\label{sec:experimentations}

We compare two prompting strategies
as depicted in Figure~\ref{fig:study_diagram}.
\textit{Zeroshot Prompting}
is when the LLM is given the commit message and the \textit{git diff}, and is asked to write the ADD. This approach evaluates the model's ability to infer architectural intent directly from commit-level artifacts and distinguish architectural changes from implementation details. 
\textit{Fewshot Prompting}
provides the LLM with examples of previously documented ADDs from the initial dataset to guide generation toward expected architectural abstraction and structure. Relevant examples are dynamically retrieved using embedding-based similarity search over previous $(\textit{commit diff}, \textit{ADD})$ pairs. During inference, 
most semantically similar ADDs are included as in-context examples, enabling the model to better align outputs with architectural descriptions.
The prompt template used is provided in our \href{https://\replpkg}{replication package}.


\subsubsection{Evaluation}

\textit{Automated metrics.}
We use established NLP similarity metrics to compare the extracted ADDs with the corresponding documented decisions. Specifically, we employ lexical metrics like ROUGE, BLEU, and METEOR, as well as the semantic metric BERTScore, which are widely used in text generation tasks \cite{guo-etal-2024-appls}.
Higher scores on these metrics indicate better alignment between extracted and documented design decisions. BERTScore typically yields higher baseline values since it measures semantic similarity rather than exact matches, so even small gains can reflect meaningful improvements in alignment.

\textit{Manual inspection.}
Automated metrics cannot tell apart real architectural reasoning from rewording of implementation details, so we complement them with a structured manual inspection. We defined a coding rubric \emph{a priori} with three dimensions, each assessed against explicit criteria: \emph{abstraction level} (whether the output describes the design decision or low-level implementation details), \emph{verbosity} (whether its length is comparable to the documented ADD, or roughly $2\times$ or more longer/shorter), and \emph{rationale} (whether it explains \emph{why} the change was made, not just \emph{what} changed). One author applied this rubric to all 60 outputs of the best-performing model (Gemini~3 Pro), recording a judgement and a short supporting note for each dimension. To limit subjectivity, the criteria were fixed before coding began and applied uniformly across all outputs.

\section{Results}
\label{sec:results}


\subsection*{RQ$_1$: Zeroshot LLM alignment with documented ADDs}

\textit{Automated metrics.}
As reported in Table~\ref{tab:results}, all four models reached a BERT-F1 above 0.81, which shows that LLMs can pick up architectural information from commits even without any task-specific help. Gemini was the best overall (BERT-F1 = 0.828, ROUGE-L = 0.134, BLEU = 0.013, METEOR = 0.175), followed by Qwen (BERT-F1 = 0.823). The word-overlap scores stayed low across all models, which means that the generated ADDs use different wording from the documented ones, even when they cover the same idea.

\textit{Manual inspection.}
The main problems observed were verbosity, wrong level of abstraction, and missing rationale. In about two thirds of the outputs, the generated ADD was 2 to 10 times longer than the documented one, as the model wrote a numbered list of small items instead of stating the decision. For example in \texttt{CASSANDRA-15505}, a one-line ADD about intercepting messages between nodes became a six-item list covering filter ordering, serialisation, and lifecycle changes. Most outputs also included at least one code-level detail (class names or configuration values). In \texttt{YARN-5889}, the documented ADD is to move user-limit computation out of the heartbeat path, but the generated ADD instead listed the classes involved (\texttt{AbstractUsersManager}, \texttt{UsersManager}). Many of the outputs also said \emph{what} changed but not \emph{why}. \texttt{CASSANDRA-3411} lists segment recycling, pre-allocation, and mmap I/O without saying these are done for commitlog performance.

\begin{table*}[htbp]
\centering

\caption{Performance comparison across the two approaches. 
Best-performing model for each metric within an approach is highlighted in bold.
}
\label{tab:results}

\begin{tabular}{l l | c c c c c }

\toprule
Approach & Model & ROUGE-L & BLEU & METEOR & BERT-F1 & Support \\
\midrule
\multirow{4}{*}{\makecell{Zeroshot}}
 & Kimi & 0.092 & 0.003 & 0.155 & 0.814 & 30 \\
 & Deepseek & 0.098 & 0.007 & 0.169 & 0.813 & 30 \\
 & Qwen & 0.109 & 0.007 & 0.170 & 0.823 & 30 \\
 & Gemini & \textbf{0.134} & \textbf{0.013} & \textbf{0.175} & \textbf{0.828} & 30 \\

\midrule

\multirow{4}{*}{\makecell{Fewshot}}
 & Kimi & 0.104 & 0.009 & 0.180 & 0.827 & 30 \\
 & Deepseek & 0.105 & 0.009 & 0.180 & 0.820 & 30 \\
 & Qwen & 0.124 & 0.009 & \textbf{0.191} & 0.829 & 30 \\
 & Gemini & \textbf{0.152} & \textbf{0.014} & 0.165 & \textbf{0.847} & 30 \\

\bottomrule
\end{tabular}
\end{table*}

\subsection*{RQ$_2$: Impact of fewshot prompting on ADD alignment}

\textit{Automated metrics.}
Fewshot prompting improved scores across all four models on most metrics, with the biggest gains on ROUGE-L and BERT-F1. Gemini remained the best overall (BERT-F1 = 0.847, ROUGE-L = 0.152, BLEU = 0.014). 
Qwen achieved the highest METEOR (0.191). The clear exception was METEOR for Gemini, which dropped from 0.175 to 0.165, suggesting that the retrieved examples sometimes push the wording further from the documented ADD even when the overall semantic match improves.

\textit{Manual inspection.}
About half of the very long zeroshot outputs were tightened to roughly the documented ADD's length. \texttt{CASSANDRA-15505} went from a six-item list to two sentences, \texttt{HADOOP-14642} from a five-item list to a single paragraph, and \texttt{YARN-5561} from a four-item REST-spec list to one sentence. The retrieved examples appear to teach the model the expected length and style of an ADD. However, implementation details still appeared in some outputs, and the missing-rationale problem was not addressed. The retrieved examples carry the documented ADDs but not the reasoning behind them, so they give the model no push to recover the \emph{why} of a decision. Fewshot prompting therefore fixed the \emph{structure} of the ADDs (length and style) but not their content, leaving the extraction of concise, well-grounded ADDs from commits an open problem.

\section{Discussion}
\label{sec:discussion}

\textbf{\textit{Key Learnings:}}
\textit{Commits show \emph{what} changed, not \emph{why}.}
The model usually named changes well but left out the rationale for them. For example, \texttt{YARN-5889} lists classes involved in the user-limit refactor but does not mention the heartbeat write-lock problem that motivated the change. 
This is a fundamental limitation of using commits as the sole input. The diff shows the change, but the reason for it usually lives in artifacts outside the commit, such as issue discussions, design documents, or code reviews. Commit messages sometimes hint at reason, but they are often short and code-focused, so the model
has little to work with.

\textit{Difficulty in maintaining the correct level of abstraction.}
Most zeroshot outputs mentioned code-level items implementations class names and configuration values. Fewshot prompting made the outputs shorter, as seen in case of \texttt{CASSANDRA-15505}, but it did not consistently help to maintain the correct architectural abstraction. The retrieved examples teach the model the expected length and style of an ADD, but not how to abstract over a diff.

\textbf{\textit{Implications for Researchers:}}
These observations suggest that ADD extraction from commits is an architectural reasoning task rather than a pure summarisation task. Since commits primarily encode implementation changes, recovering the rationale likely requires combining commits with other artifacts 
where the reasoning is more explicit \cite{soliman2016exploring,Bhat}. 
The results also reveal the limitations of automated similarity metrics, which cannot reliably distinguish architectural reasoning from implementation-level paraphrasing, highlighting the need for evaluation methods that better capture architectural meaning.

\textbf{\textit{Implications for Practitioners:}}
LLM-generated ADDs can be useful as a first draft for AKM, which can lower the documentation effort significantly. However, the output should be treated as a starting point and not as a finished ADD. The rationale for the decision usually needs to be added or checked, and implementation details should be removed.

\section{Threats to Validity}
\label{sec:threats}

\textit{Construct validity:} The automated metrics (ROUGE-L, BLEU, METEOR, and BERTScore) may not capture architectural meaning, since the same ADD can be written with different words and at different levels of detail. 
We reduce this risk by combining all four metrics with a manual review. The manual inspection was done by a single author using free-text notes rather than a formal coding process with multiple reviewers, which can introduce subjectivity. We mitigate this by fixing the coding criteria before inspection and applying them uniformly to all outputs.

\textit{Internal validity:} LLM outputs depend on how the prompt is worded. We reduce this risk by using the same prompt template for all runs. There is also a risk of data leakage, because the Apache projects we use are public and the corresponding commits and ADDs may appear in the LLMs' pretraining data. We reduce this risk by combining the automated metrics with manual inspection focused on meaning rather than exact wording. 

\textit{External validity:} The dataset covers 30 ADDs from three Apache repositories (Cassandra, Hadoop and Tajo) and is biased towards Java-based systems. ADDs were picked for being clearly written and tied to a single commit, which introduces selection bias. Manual inspection covers only Gemini outputs so problems seen on weaker models may differ. The findings should be read as preliminary, and larger studies across more projects, languages, and models are needed.

\section{Conclusion and Future Directions}
\label{sec:conclusion}

We presented a preliminary exploratory study on extracting ADDs from source code commits using LLMs. LLMs can recover architectural information from commits to a moderate extent (BERT-F1 above 0.81 for all four models), and fewshot prompting improves the scores. However, the generated ADDs are often too implementation-focused and leave out the rationale behind the change. This suggests that extracting ADDs from commits is not purely a summarisation task, but a higher-level architectural reasoning problem. 
These limitations stem from the commit input itself and will likely persist as models evolve.

For future work, we plan to add other sources such as issue discussions, code reviews and repository-level context to help the model find the rationale better. Larger-scale evaluations and improved evaluation methodologies are needed to better assess architectural alignment and generation quality.


%
%
\bibliographystyle{abbrv}
\bibliography{references}

\end{document}